\documentclass[10pt]{iopart}  
\usepackage{siunitx}
\usepackage{graphicx}
\usepackage{cite}
\usepackage{braket}

\begin{document}

\title{Assembled arrays of Rydberg-interacting atoms}
\author{Malte Schlosser, Daniel Ohl de Mello, Dominik Sch\"{a}ffner, Tilman Preuschoff, Lars Kohfahl and Gerhard Birkl}
\address{Institut f\"{u}r Angewandte Physik, Technische Universit\"{a}t Darmstadt, Schlossgartenstra\ss e 7, 64289 Darmstadt, Germany}
\ead{apqpub@physik.tu-darmstadt.de}
\vspace{10pt}
\begin{indented}
\item[]February 2020
\end{indented}

\begin{abstract}
Assembled arrays of individual atoms with Rydberg-mediated interactions provide a powerful platform for the simulation of many-body spin Hamiltonians as well as the implementation of universal gate-based quantum information processing. We demonstrate the first realization of Rydberg excitations and controlled interactions in microlens-generated multisite trap arrays of reconfigurable geometry. We utilize atom-by-atom assembly for the deterministic preparation of pre-defined 2D structures of rubidium Rydberg atoms with exactly known mutual separations and selectable interaction strength. By adapting the geometry and the addressed Rydberg state, a parameter regime spanning from weak interactions to strong coupling can be accessed. We characterize the simultaneous coherent excitation of non-interacting atom clusters for the state $\mathrm{57D_{5/2}}$ and analyze the experimental parameters and limitations. For configurations optimized for Rydberg blockade utilizing the state $\mathrm{87D_{5/2}}$, we observe collectively enhanced Rabi oscillations.
\end{abstract}

\vspace{2pc}
\textbf{Original Citation:}\\\indent
\textbf{J. Phys. B: At. Mol. Opt. Phys. 53, 144001 (2020)}\\\indent
\textbf{DOI: 10.1088/1361-6455/ab8b46}\\\indent
(Contribution to J. Phys. B - Special Issue on Interacting Rydberg Atoms)\\\indent

\ioptwocol

\section{Introduction}

Individual atoms in arrays of optical traps constitute a versatile platform of experimentally well accessible quantum systems for quantum simulation and computation \cite{Saffman2010,Saffman2016,Browaeys2016,Gross2017,Schauss2018,Shaffer2019,Browaeys2020}. Their multisite geometry facilitates the allocation of large-scale quantum resources \cite{OhldeMello2019,Wang2020,Schlosser2020Talbot} while being configurable \cite{Barredo2014,Kim2018,Kim2019,Schaffner2020} in order to represent a broad range of model systems. Atom-by-atom assembly of the target patterns provides the crucial ability to deterministically prepare the system of interest in a defect-free manner \cite{Barredo2016,Endres2016,Kim2016,Barredo2018,OhldeMello2019,Brown2019}. Interactions of variable strength and range can be introduced on demand by the site-resolved excitation of Rydberg states \cite{Barredo2014,Zeiher2015,Labuhn2016,Lee2019,Madjarov2020,Wilson2019}. High-fidelity coherent control of the dynamics has led to the creation of entanglement \cite{Wilk2010,Picken2018,Omran2019,Jo2020}, the implementation of quantum gates \cite{Isenhower2010,Zeng2017,Levine2019}, and the simulation of quantum Ising models and XY magnets including the study of topological phases \cite{Kim2018,Lienhard2018,Leseleuc2018a,Keesling2019}. At the same time, these systems hold potential for the parallelization of universal gate operations en route to large-scale quantum computing \cite{Saffman2010,Saffman2016,Levine2019,Graham2019}.\\
In this work, we expand Rydberg many-body physics to our scalable micro-optical platform \cite{Birkl2000,Dumke2002,Birkl2007,Schlosser2011,OhldeMello2019}. We demonstrate significant progress for microlens-generated trap arrays facilitating the deterministic preparation of 2D Rydberg-coupled systems of tunable interactions and adaptable geometry. The implementation of defect-free and controllable ordered configurations with defined atom positions enables the time- and site-resolved observation of coherent dynamics in the Rydberg-blockaded regime.

\section{Methods}
\subsection*{Microlens-generated arrays of single atoms}

The illumination of a microlens array (MLA) with laser light of suitable wavelength creates a multitude of focused-beam dipole traps in a parallelized fashion \cite{Dumke2002,Schlosser2011}. The trap positions reflect the arrangement of lenslets on the micro-fabricated optical element where large-scale periodic quadratic and triangular as well as custom geometries are available. On top of this, rapid prototyping of user-defined trap patterns has been implemented by 3D printing of micro-optics using direct laser writing techniques \cite{Schaffner2020}. Configurations of interleaved trap arrays add further flexibility by enabling the parallelized dynamic positioning and transport of stored-atom quantum systems \cite{Lengwenus2010,Schlosser2020Talbot}. The number of traps with sufficient depth is determined by the number of lenslets that are illuminated with sufficient power. As lithographically produced MLAs may contain millions of lenslets, the number of traps in current experiments is only technically limited by the available laser power.
\begin{figure*}
	\centering
	\includegraphics[width=1\linewidth]{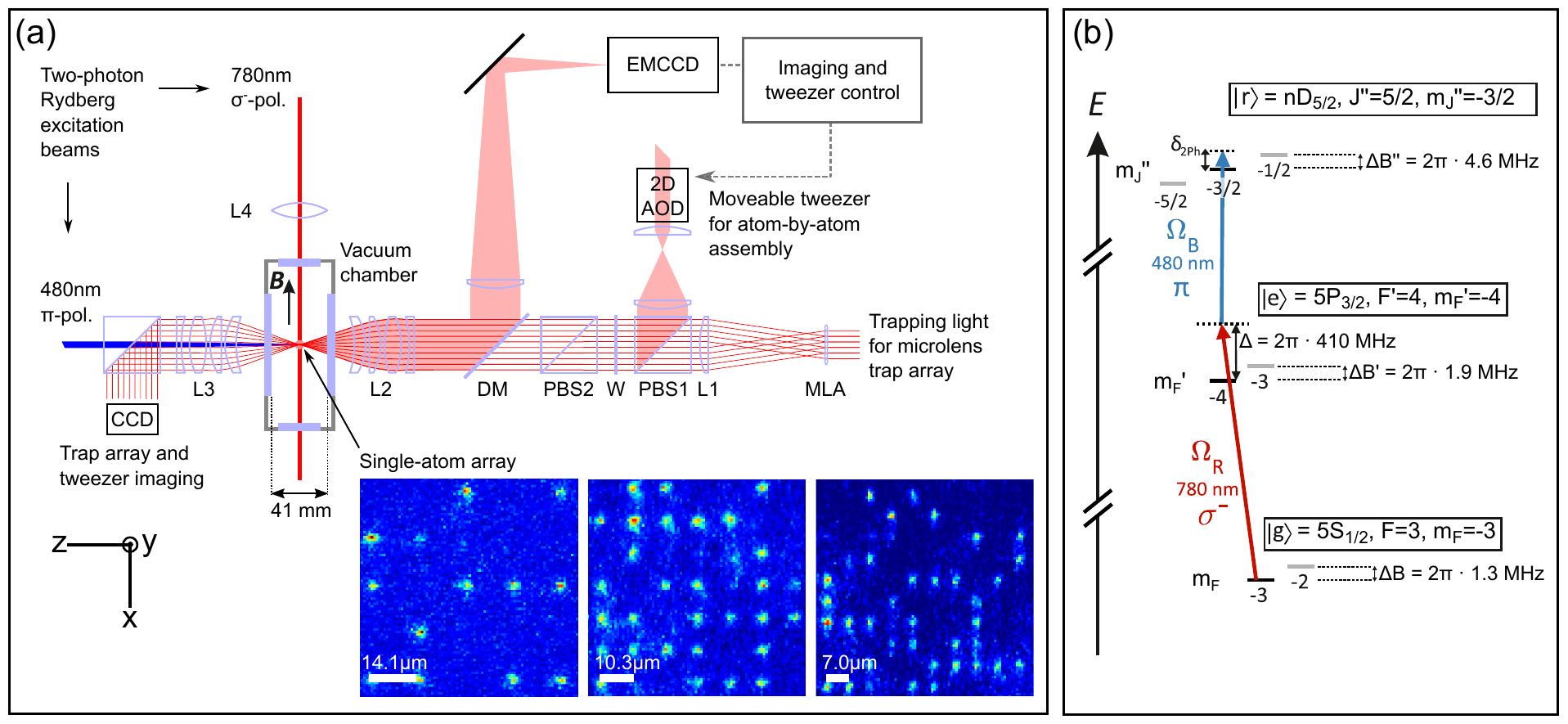}
	\caption{(a) Microlens-generated arrays of single atoms.
		A top view schematic of the experimental setup and the relevant beam paths is shown. The incident dipole trap beam passes a quadratic-grid microlens array (MLA). The resulting focal spot array is reimaged into a vacuum chamber to generate the trap array by demagnifying relay optics of lens L1 and microscope objective L2 (numerical aperture $\mathrm{NA=\SI{0.25(2)}{}}$).
		This same objective is used in conjunction with a dichroic mirror DM to image the fluorescence light of trapped atoms onto an EMCCD camera.
		Captured single shot images of individual atoms stored in trap arrays with pitch of \SI{14.1}{\micro\meter}, \SI{10.3}{\micro\meter}, and \SI{7.0}{\micro\meter} are displayed on the bottom.
		An independent beam of trapping light directed through a 2D acousto-optic deflector (AOD) creates a moveable optical tweezer. Its beam path is superposed with the trap array light on a polarizing beam splitter PBS1 and the polarizations are aligned using half wave plate W and PBS2.
		After passing the vacuum chamber, the light fields of trap array and tweezer are imaged onto a CCD camera for analysis and optimization using a second microscope objective L3.		
		The laser beams for two-photon Rydberg excitation of \SI{480}{\nano\meter} (\SI{780}{\nano\meter}) are linearly (circularly) polarized	and address the trap array through L3 (L4) on orthogonal beam paths. A magnetic field $\mathbf{B}$ is applied along the x direction to create a quantization axis and lift the degeneracy of Zeeman levels. (b) Two-photon scheme for excitation to Rydberg states. Atoms are initially prepared in the $\ket{g} = \ket{F=3,\,m_F=-3}$ ground state. The combination of $\sigma^-$- and $\pi$-polarized light excites the Rydberg state $\ket{r} = nD_{5/2}$ with magnetic quantum number $m_J^"=-3/2$.
	}
	\label{fig:1}
\end{figure*}
\\Figure~\ref{fig:1}\,(a) depicts the current experimental platform of our rubidium atom array. We use MLAs of plano-convex lenslets with spherical curvature in a quadratic grid. As visualized in the schematic, the MLA is illuminated with a Gaussian beam of linearly polarized trapping light produced by a titanium-sapphire laser. The typical trap depth is ${U_0\approx k_B\cdot\SI{1}{\milli\kelvin}}$, which requires about \SI{1}{mW} of trapping light per site at a typical wavelength of \SI{798}{\nano\meter}, and results in a maximum number of trapping sites on the order of 2500 \cite{OhldeMello2019} in the focal plane. The spot pattern of the focal plane is reimaged into the vacuum chamber using demagnifying relay optics consisting of an achromatic lens (L1, focal length $f_1$) and a microscope objective (L2) of \SI{37.5(10)}{\milli\meter} effective focal length and numerical aperture of $\mathrm{NA=\SI{0.25(2)}{}}$. This creates an array of diffraction limited spots with waists of \SI{1.45(10)}{\micro\meter}. Geometry and pitch are adaptable by either changing the MLA or the focal length $f_1$. Table~\ref{tab:1} lists the parameters of three trap arrays utilized in this work.\\
In order to uniformly fill the array, we superpose a cloud of ${}^{85}\mathrm{Rb}$ atoms that is collected and cooled in a six-beam magneto-optical trap (MOT) at a detuning of $\SI{-10}{\mega\hertz}$ relative to the $|\mathrm{5S_{1/2}, F=3\rangle\leftrightarrow|5P_{3/2}, F'=4}\rangle$ transition and subsequently expanded during a short time-of-flight phase. The atoms are loaded into the traps during a $\SI{30}{\milli\second}$ stage of polarization-gradient cooling and light assisted collisions at a detuning of \SI{-52}{\mega\hertz}. The collisional blockade effect \cite{SchlosserN2001,Gruenzweig2011,Brown2019} causes a trap occupation of either zero or one atom. The single atoms are probabilistically distributed in the array. We observe an upper bound of $\SI{55}{\%}$ for this initial filling fraction.\\
Out of the multilayer trapping geometry created by the Talbot effect \cite{Schlosser2020Talbot}, for this work we restrict our system to one single layer of atoms by removing all atoms from the layers outside the focal plane by radiation pressure. For this purpose, we apply a beam of laser light resonant to the $|\mathrm{5S_{1/2}, F=3\rangle\leftrightarrow|5P_{3/2}, F'=4}\rangle$ transition perpendicular to the dipole trap axis with a shadow along the focal plane.\\
We use an electron-multiplying camera (EMCCD) for site-resolved atom detection. Atom fluorescence is induced by the cooling light and collected by the microscope objective (L2). The fluorescence is separated from the trapping light by a dichroic mirror (DM) and imaged with a magnification of $20$, ensuring a clear spatially resolved detection of the atoms. We obtain a single atom detection fidelity larger than \SI{99}{\%} for every array site using an integration time of \SIrange[range-phrase = {\text{~to~}} ,range-units=single]{50}{70}{\milli\second}. Figure~\ref{fig:1}\,(a) shows single shot images of trapped atoms in a selection of microlens-generated arrays where we have varied the trap period (pitch) between \SI{14.1}{\micro\meter} and \SI{7.0}{\micro\meter}. Additional parameters are listed in Tab.~\ref{tab:1}.
\begin{table}
	\caption{\label{tab:1}
		Parameters of microlens arrays (MLA) and trap arrays.
		All MLAs have a quadratic grid geometry.}
	\footnotesize
	\centering{
		\begin{tabular}{@{}c|cc|c|cc}
			\br
			Par.&\multicolumn{2}{c|}{Microlens array}&L1&\multicolumn{2}{c}{Trap array}\\
			set&\textit{pitch}&\textit{NA}&$f_1$&\textit{pitch}&$\lambda$\\
			\mr
			1&\SI{30}{\micro\meter}&0.14&\SI{80}{\milli\meter}&\SI{14.1(4)}{\micro\meter}&\SI{798.6}{\nano\meter}\\
			2&\SI{110}{\micro\meter}&0.03&\SI{400}{\milli\meter}&\SI{10.3(3)}{\micro\meter}&\SI{797.3}{\nano\meter}\\
			3&\SI{75}{\micro\meter}&0.03&\SI{400}{\milli\meter}&\SI{7.0(2)}{\micro\meter}&\SI{797.3}{\nano\meter}\\
			\br
		\end{tabular}\\}
	Parameters common to all trap arrays:\\
	trap waist ($1/e^2$-radius) $w_0=\SI{1.45(10)}{\micro\meter}$\\
	trap depth $U_0\approx k_B\cdot\SI{1}{\milli\kelvin}$
\end{table}
\normalsize\\
\subsection*{Atom-by-atom assembly}

\begin{figure}
	\includegraphics[width=\linewidth]{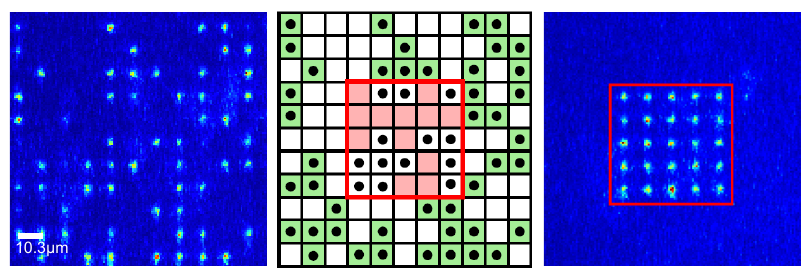}
	\caption{
		Defect-free assembly of a 2D cluster of individual atoms.
		The initial distribution of individual atoms in the array (left) is analyzed and converted to a boolean occupation grid (middle). Occupied sites are marked with a black dot. All sites are classified as correctly occupied (white), reservoir atom (green), and empty target site (red). A shortest-move heuristic is used to compute the required rearrangement sequence. Upon successful execution of the atom moves, defect-free filling of the target structure (red box) is observed (right).
	}
	\label{fig:2}
\end{figure}

\begin{figure}
	\includegraphics[width=\linewidth]{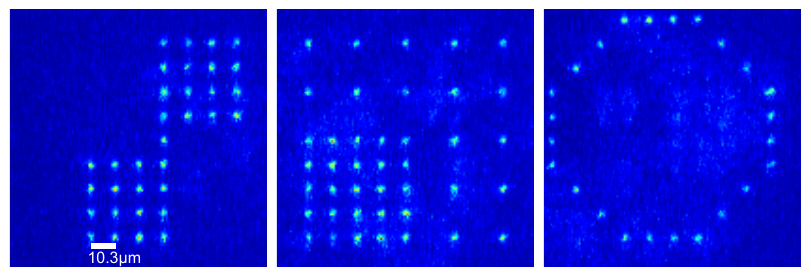}
	\caption{
		Gallery of assembled atom patterns (see text for details). The underlying array is of parameter set 2 (Tab.~\ref{tab:1}).
	}
	\label{fig:3}
\end{figure}

The deterministic assembly of a target structure is visualized in Fig.~\ref{fig:2}. We use a path finding algorithm to calculate a sequence of atom rearrangements from the detected initial distribution of atoms (left). The source array consists of up to ${19{\times}19}$ traps. Within this grid, all occupied traps other than the centered target structure (red box) serve as reservoir (middle). The moveable optical tweezer carries out the rearrangement sequence in order to achieve a defect-free pattern (right). We employ a greedy algorithm which pairs each vacant target site with the closest available reservoir atom, starting in the center of the target structure. Atoms are moved along the virtual grid lines connecting the sites. If a calculated path contains an occupied trap, this obstacle atom is moved into the target trap instead and the original reservoir atom takes its place. The algorithm optimizes the rearrangement sequence by choosing the path with the fewest obstacles for every single-atom move. This approach ensures a time-efficient solution. It shows a polynomial scaling with the number of vacant target sites and generates a not necessarily optimal, but sufficient set of single-atom moves.\\
The tweezer position can be dynamically controlled using a 2D acousto-optic deflector system and signal generation by FPGA-controlled direct digital frequency synthesis. 
As depicted in Fig.~\ref{fig:1}\,(a), the tweezer beam path is superposed onto the trap array beam on a beam splitter. The light fields are of near-identical wavelength with a frequency offset of several hundred megahertz. After being focused into the vacuum chamber by the microscope objective (L2), the spot size is \SI{2.0(1)}{\micro\meter} and the potential depth evaluates to ${k_B\cdot\SI{0.5}{\milli\kelvin}}$. The addressable region exceeds ${\SI{400}{\micro\metre}\times\SI{400}{\micro\metre}}$.\\
The elementary rearrangement operation between two array sites is implemented by (1) setting the tweezer's depth to zero and superposing its position with the source site, (2) capturing the atom in the tweezer by ramping the tweezer depth to maximum, (3) moving the tweezer to the target site, and (4) transferring the atom to the target trap by ramping the tweezer depth to zero again. For the data presented in this paper, we linearly ramp the depth with a typical ramp duration of \SI{200}{\micro\second} and use sine-shaped transport trajectories. A typical duration for a single transport is \SI{1}{\milli\second}, which is well within the limits of adiabaticity \cite{Schlosser2012}. The full rearrangement sequence consists of (1) lowering the array depth by typically a factor of five, (2) applying the required set of atom moves, (3) restoring the original array depth, and (4) detecting the resulting atom positions. This amounts to a total duration of \SIrange[range-phrase = {\text{~to~}} ,range-units=single]{60}{120}{\milli\second}, depending on the number of atom moves. We perform up to 15 rearrangement sequences in rapid succession in order to increase achievable structure sizes and success probabilities \cite{OhldeMello2019}.\\
Intentionally empty target sites in the assembled patterns are realized by either transporting surplus atoms to outlying sites within the rearrangement sequence or by removal of these atoms from the array upon successful assembly. The gallery of Fig.~\ref{fig:3} demonstrates the capabilities of the implemented atom-by-atom assembler. Within the trap array of parameter set 2 (Tab.~\ref{tab:1}), the pattern of an atomtronic diode \cite{Sturm2017} (left), a supergrid of occupied sites of the underlying periodic array (middle) and a 1D system of atoms with periodic boundary conditions (right) are created.
After generating the targeted cluster, the tweezer can be used to redistribute atoms between different target structures \cite{OhldeMello2019} also preserving quantum state coherence, as we have demonstrated in \cite{Lengwenus2010}.
\subsection*{Two-photon excitation of Rydberg states}
\begin{figure*}[t]
	\centering
	\includegraphics[width=0.9\linewidth]{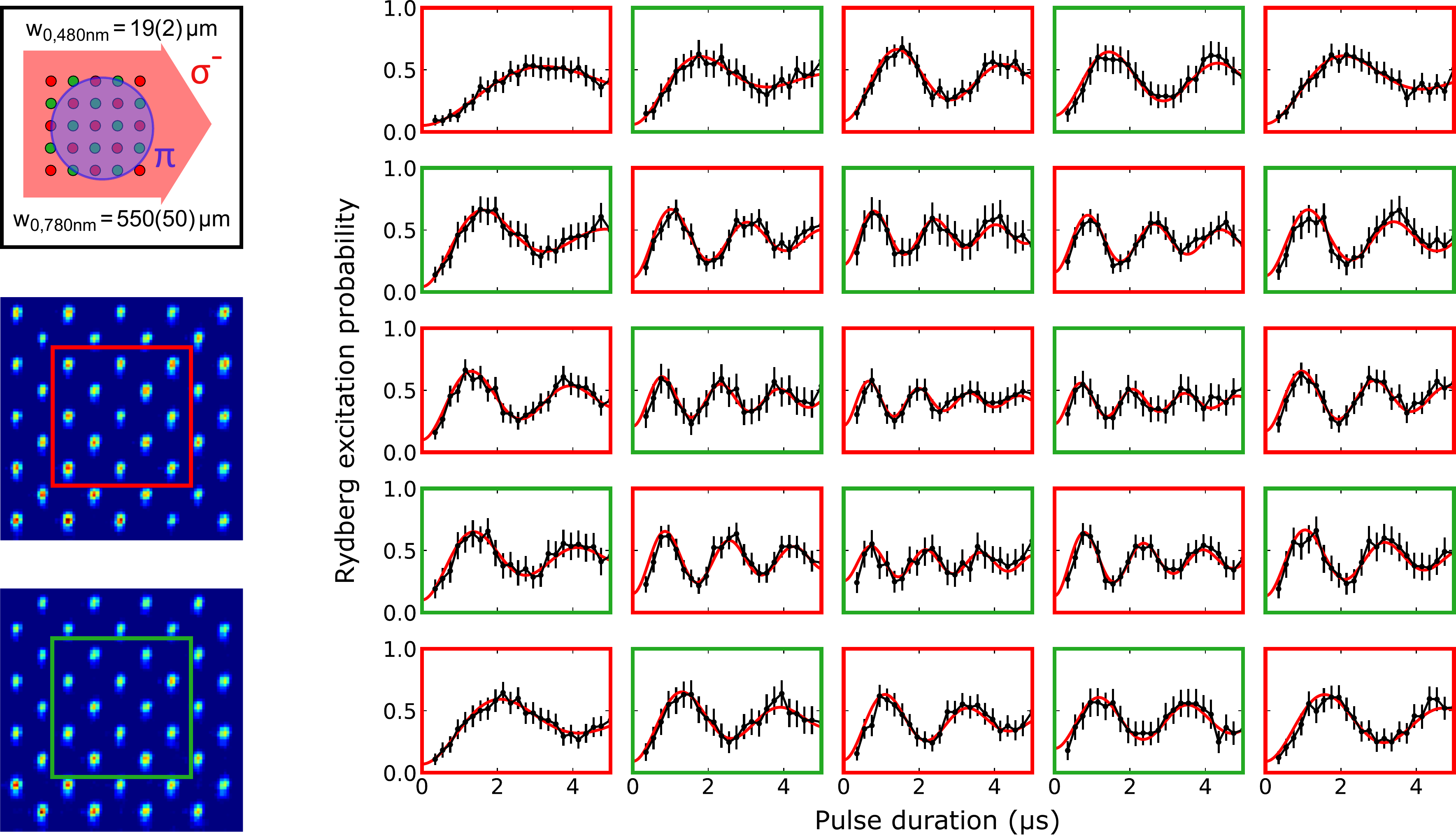}
	\caption{
		Simultaneous Rabi oscillations in the \SI{7.0}{\micro\meter} pitch array (parameter set 3 of Tab.~\ref{tab:1}).
		(Left) The single atom array is coupled to the Rydberg state $\mathrm{|57D_{5/2},m_J=-3/2\rangle}$ using a two-photon $\sigma$-$\pi$ scheme.		
		In two separate measurements, the atoms are assembled in a checkerboard pattern (averaged images) to avoid nearest neighbor interactions when excited to the Rydberg state, which would increase dephasing. 
		The data of the complementary patterns (red and green frames) is combined for analysis (right).
		Error bars correspond to statistical uncertainties.
	}
	\label{fig:4}
\end{figure*}
We use the two-photon excitation scheme depicted in Fig.~\ref{fig:1}\,(b) to coherently couple the ground-state $\mathrm{^{85}Rb}$ atoms via the D2-line to Rydberg states. The setup consists of an interference-filter-stabilized external cavity diode laser (ECDL) at \SI{780}{\nano\meter} and a grating-stabilized ECDL at \SI{960}{\nano\meter} whose output is amplified and frequency doubled to \SI{480}{\nano\meter}. Both lasers are frequency-locked to an ultra-stable reference cavity using the Pound-Drever-Hall technique. The cavity finesse was determined to ${\mathcal{F}_{780}=\SI{31500(100)}{}}$ and ${\mathcal{F}_{960}=\SI{46700(120)}{}}$. We obtain an upper bound on the laser linewidths of \SI{11}{\kilo\hertz} (\SI{4.5}{\kilo\hertz}) for the \SI{780}{\nano\meter} (\SI{960}{\nano\meter}) laser on a timescale of \SI{1}{\milli\second} from analyzing the light transmitted through the cavity. The cavity is housed in a vacuum chamber and operated at its zero crossing of thermal expansion. This strongly decouples the system from drifts in air pressure and ambient temperature. The cavity resonance is monitored by tracking the frequency beat of the \SI{780}{\nano\meter} Rydberg laser and a reference laser stabilized to a rubidium transition \cite{Preuschoff2018}. The observed drift is in accordance with the predicted aging of the spacer material. Acousto-optic modulators (AOM) are used to offset the laser frequency with regard to the cavity modes. The detuning from the intermediate state $\mathrm{|5P_{3/2}, F'=4}\rangle$ is set to \SI{410(10)}{\mega\hertz}.\\
In Fig.~\ref{fig:1}\,(a) the beam paths of the Rydberg excitation lasers on the experimental table are depicted. The beams excite the atoms into the Rydberg state ${\ket{r} = nD_{5/2}}$ with magnetic quantum number ${m_J^"=-3/2}$ via a $\sigma^-$-$\pi$ polarization scheme, as is illustrated in Fig.~\ref{fig:1}\,(b). To ensure a well-defined quantization axis, a magnetic field (${|B|=\SI{2.75}{G}}$) is applied along the x-axis, which lifts the degeneracy of the magnetic sublevels. The resulting splitting of Zeeman levels with $\Delta m = 1$ is $\Delta B = 2\pi\cdot\SI{1.3}{\mega\hertz}$ for the ground state, $\Delta B^\prime = 2\pi\cdot\SI{1.9}{\mega\hertz}$ for the intermediate state and $\Delta B^{\prime\prime} = 2\pi\cdot\SI{4.6}{\mega\hertz}$ for the Rydberg state. The \SI{780}{\nano\meter} Rydberg beam is circularly polarized and has a waist ($\mathrm{1/e^2}$) of \SI{550(50)}{\micro\meter} and a power of up to \SI{760}{\micro\watt} at the position of the atoms. The \SI{480}{\nano\meter} beam is linearly polarized, illuminating the atoms with an estimated optical power of \SI{45}{\milli\watt} and a beam waist of \SI{19(2)}{\micro\meter} (see Sec.~\ref{sec:results}). Both Rydberg lasers are stabilized in intensity via a sample-and-hold scheme to ensure pulses with constant laser power over the course of a measurement. In a typical experimental sequence, we initially prepare all atoms in the $\mathrm{|5S_{1/2}, F=3, m_F=-3}\rangle$ state. For this purpose, the trap depth is ramped down by a factor of five and $\sigma^-$-polarized laser light near-resonant to the $|\mathrm{5S_{1/2}, F=3\rangle\leftrightarrow|5P_{3/2}, F'=3}\rangle$ transition is applied. The state preparation beam is superposed with the beam of the \SI{780}{\nano\meter} Rydberg laser. Directly afterwards, the traps are switched off for \SI{10}{\micro\second} and the two-photon Rydberg excitation light is applied. The effective pulse length is controlled through an AOM in the \SI{780}{\nano\meter} Rydberg beam with switching times on the order of \SI{50}{\nano\second}. While ground-state atoms are recaptured into the traps, Rydberg atoms are lost, allowing for state-dependent detection.

\section{Results}
\label{sec:results}
\subsection*{Spatially resolved coherent ground-to-Rydberg-state dynamics in assembled atom arrays}

Applying the Rydberg excitation light at the two-photon resonance with increasing pulse length causes the system to coherently oscillate between the ground and Rydberg state. These Rabi oscillations are shown in Fig.~\ref{fig:4} for the Rydberg state $\mathrm{|57D_{5/2},m_J=-3/2\rangle}$.\\
The trap array is of parameter set 3 (Tab.~\ref{tab:1}) with a separation of neighboring sites of \SI{7.0}{\micro\meter}. In order to reduce the van der Waals interaction between neighboring Rydberg atoms in the array and thus isolate the undisturbed Rabi frequencies, atom assembly is utilized to prepare a supergrid of atoms in a checkerboard pattern. The pattern is inverted in a second measurement. Averaged fluorescence images of both patterns are shown in Fig.~\ref{fig:4} (left). The Rydberg excitation probability is extracted from analyzing the atom recapture rate after the excitation pulse. Figure~\ref{fig:4} (right) displays the merged data from both patterns as a function of pulse length, leading to the observation of Rabi oscillations with varying amplitude and frequency across the array. Each data point is based on an average number of 70 experimental realizations where an atom was initially present, with error bars corresponding to a statistical uncertainty.\\
The red solid lines are fits to the data based on a simple model obtained from solving the optical Bloch equations for the effective two-level system including spontaneous emission from the intermediate state. We extract Rabi frequencies in the range of ${\Omega/(2\pi)=\SIrange[range-phrase = {\text{~to~}} ,range-units=single]{0.15}{0.75}{\mega\hertz}}$. The average damping constant is ${\gamma=\SI{0.37(9)}{\micro\second^{-1}}}$. As the \SI{780}{\nano\meter} Rydberg beam can be assumed to be homogeneous within the analyzed area, information about the position and spatial profile of the \SI{480}{\nano\meter} Rydberg laser beam is obtained from the distribution of Rabi frequencies across the array. A fit to $\Omega^2(r)$ yields a beam waist of \SI{19(2)}{\micro\meter} and a maximum Rabi frequency of ${\Omega/(2\pi)=\SI{0.77(2)}{\mega\hertz}}$ at the beam center.

\subsection*{Rydberg blockade and collective enhancement of Rabi frequencies}
\begin{figure}
	\includegraphics[width=\linewidth]{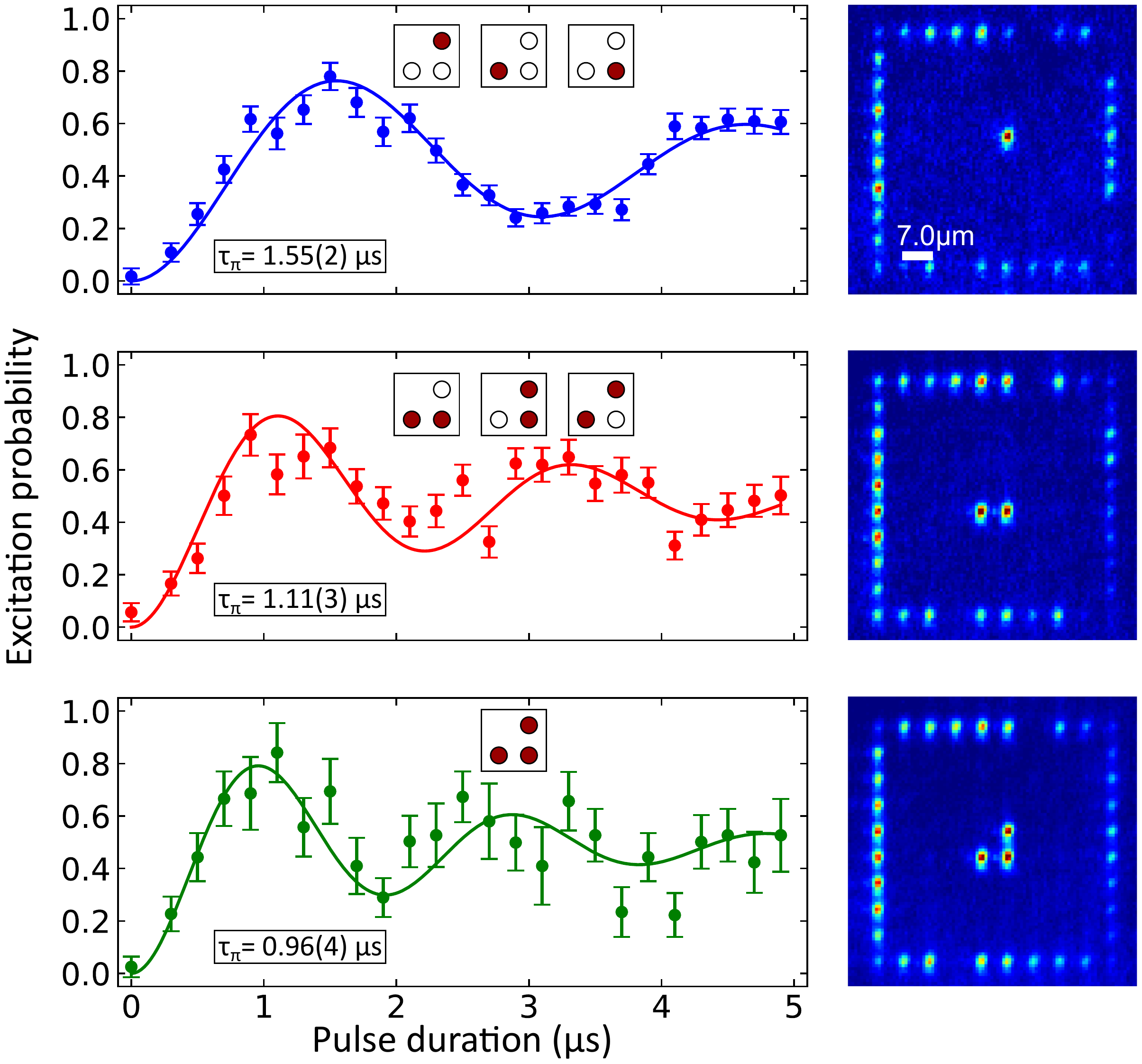}
	\caption{
		Collective enhancement of the Rabi frequency through Rydberg blockade. The ground state atoms are coupled to the Rydberg state $\mathrm{|87D_{5/2},m_J=-3/2\rangle}$ and the pulse length of the excitation light is varied. The graphs depict the probability of observing a single Rydberg excitation in a system of one, two and three atoms, respectively (top to bottom).
		Error bars correspond to statistical uncertainties and the data are corrected for static preparation and detection errors. The observed damping of the oscillations is due to a combination of effects such as spontaneous decay, Doppler broadening, and position fluctuations of the atoms (see Sec.~\ref{sec:discussion} for details). The insets indicate the initially prepared atom structures contributing to the signal. Exemplary fluorescence images are shown on the right. Atom rearrangement was applied and the data were post-selected to ensure exclusive occupation of the central target structure within an otherwise empty ${8{\times}8}$-site region.
	}
	\label{fig:5}
\end{figure}
While the above characterization of the coherent ground-to-Rydberg-state coupling relied on measures to decouple the dynamics of atoms in different sites, we can easily adapt the experimental parameters to incorporate strong interactions within the array. Large interaction energy shifts ${|\Delta E|=|-C_6/R^6|\gg\hbar\Omega}$ result in the blockade of excitation of more than one atom inside a blockade radius ${R_b=\sqrt[6]{C_6/(\hbar\Omega)}}$, where $C_6$ is the van der Waals coefficient. For these systems, Rydberg blockade results in a collective enhancement of the Rabi frequency, scaling with the atom number $N$ as ${\Omega_N=\sqrt{N}\Omega}$ where ${\Omega\equiv\Omega_{N=1}}$ \cite{Urban2009,Gaetan2009}.\\
In the set of experiments presented in Fig.~\ref{fig:5}, assembled atom structures within this strongly interacting regime are realized. The targeted Rydberg state is $\mathrm{|87D_{5/2},m_J=-3/2\rangle}$ and the site separation is \SI{7.0}{\micro\meter}. As the orientation of interatomic axes in the array varies relative to the quantization axis, we have to take the anisotropic interaction potential for $\mathrm{D}$ states into account. Limiting the considerations to direct neighbors and diagonal neighbors, the calculated blockade energies are in the range of $|{\Delta E|/h = \SIrange[range-phrase = {\text{~to~}} ,range-units=single]{5}{24}{\mega\hertz}}$ \cite{Sibalic2017,Weber2017}. In order to minimize damping due to beating of incommensurate Rabi frequencies within the blockaded ensemble, the \SI{480}{\nano\meter} Rydberg beam was aligned such that a comparable single-atom Rabi frequency of ${\Omega/(2\pi)=\SI{0.32(3)}{\mega\hertz}}$ could be detected for a cluster of three traps without neighboring atoms. Figure~\ref{fig:5} (left) shows Rabi oscillations in the probability of detecting a single Rydberg excitation in the prepared structures. From top to bottom, the number of atoms in the initially prepared configuration was increased from one, to two, to three, with the corresponding structures indicated in the insets. The data are corrected for static preparation and detection errors (see Sec.~\ref{sec:discussion}). Solid lines correspond to fits to the data using the same model as above. The extracted values for the collective Rabi frequencies and damping constants are listed in Tab.~\ref{tab:2}. The predicted scaling of the Rabi frequency with the initial number of atoms is clearly confirmed.\\
Exemplary fluorescence images of the structures are shown in Fig.~\ref{fig:5} (right). In order to rule out any residual interactions from atoms outside the target structure, a region spanning three trap pitches in each direction of the target cluster was emptied prior to the application of the Rydberg excitation pulses. The closest distance to the atom frame around the target cluster is \SI{28}{\micro\meter}, corresponding to an interaction strength below ${|\Delta E|/h =\SI{30}{\kilo\hertz}}$. This is less than the transform-limited linewidth of the excitation, even for the longest occurring pulse durations of \SI{5}{\micro\second}. In addition to the near-deterministic preparation of the structures, the data were post-selected to only include measurements in which the depicted initial configurations including the empty surrounding area were achieved. Atoms in the surrounding frame and beyond served as a reservoir for reloading the structure after a Rydberg excitation cycle. This allowed for up to five repeated reconstruction and Rydberg excitation cycles within one experimental cycle of MOT loading.
\begin{table}
	\caption{\label{tab:2}
		Collective Rabi frequencies and damping constants.}
	\footnotesize
	\centering{
		\begin{tabular}{@{}c|cc|c}
			\br
			Atoms&\multicolumn{2}{c|}{Measured values}	&Collective scaling\\
			$N$       		&$\Omega_N/(2\pi)$ (MHz)						&$\gamma$ (\SI{}{\micro\second^{-1}})	&$\Omega_N/(\sqrt{N}\Omega)$\\
			\mr
			1                           &\SI{0.33(2)}{}   &\SI{0.32(5)}{}	&\SI{1}{}\\
			2                           &\SI{0.45(3)}{}   &\SI{0.54(11)}{}	&\SI{0.97(9)}{}\\
			3                           &\SI{0.52(5)}{}   &\SI{0.66(16)}{}	&\SI{0.92(11)}{}\\
			\br
	\end{tabular}}
\end{table}
\normalsize
\section{Discussion}
\label{sec:discussion}
\subsection*{Simultaneous Rydberg dynamics}

Simultaneous excitation in a 5x5 site region of the array could be realized and studied effectively, allowing for an accurate mapping of the individual Rabi frequencies and light shifts (Fig.~\ref{fig:4}). The observed damping and reduced contrast of the oscillation amplitude point to fast decoherence and dephasing of the dynamics as well as to imperfect excitation and detection of Rydberg states \cite{Leseleuc2018b,Levine2018}.\\
From a comprehensive analysis of our experimental parameters and limitations, we have identified the main sources of error and their contributions to the observed dynamics. The main static effect causing a reduced initial value of the observed Rydberg excitation probability is the false negative detection error of ${\SI{0.19}{}~(n=\SI{57}{})}$. This corresponds to a reduction of the observed probability of Rydberg excitation to \SI{81}{\%}, even for a factually achieved probability of \SI{100}{\%}. This value has been obtained by a simulation of the Rydberg atom recapture probability. This takes into account the thermal motion corresponding to the measured temperature of \SI{52(1)}{\micro\kelvin}, the ponderomotive potential (${\lambda=\SI{797.3}{\nano\meter}}$) and the decay rate to the ground state of \SI{0.005}{\per\micro\second}. Additional contributions stem from the state preparation fidelity of \SI{0.96}{} and a general atom loss of \SI{0.05}{} for the whole sequence associated with the truncation of the Boltzmann distribution due to decreasing the trap depth as well as losses during fluorescence imaging.\\
A noticeable damping of the oscillations is expected due to the relatively small detuning from the intermediate state. The calculated scattering rate is \SI{0.08}{\per\micro\second}, with each scattering event leading to instantaneous loss of coherence. The dominant contributors to dephasing are the Doppler effect and laser power fluctuations: The thermal distribution of atom velocities gives rise to a Doppler-broadened distribution of the effective resonance frequency of the two-photon excitation with a width of ${2\pi\cdot\SI{170}{\kilo\hertz}}$. This causes a variation of the Rabi frequency $\Omega$ for diffent atoms and increased dephasing. The power fluctuations of the Rydberg excitation light are estimated to ${\SI{5}{\%}~(\SI{480}{\nano\meter})}$ and ${\SI{2}{\%}~(\SI{780}{\nano\meter})}$. This results in a variation of the Rabi frequency $\Omega$ with an uncertainty of ${\Delta\Omega/(2\pi)=\SI{37}{\kilo\hertz}}$ and a possible modification of the effective resonance frequency of \SI{24}{\kilo\hertz}, arising from the intensity-dependent Stark shift. While the estimated position fluctuations of the atoms in the trap potentials are negligible in radial direction (${\Delta x = \Delta y = \SI{0.3}{\micro\meter}}$) an axial variation of ${\Delta z = \SI{2.4}{\micro\meter}}$ leads to a fluctuation of interaction strengths, contributing to the damping of Rabi oscillations as well.\\
While the above effects lead to simulated signals in good agreement with the experimental ones, residual discrepancies remain: The simulation underestimates damping and loss of contrast for central sites. A natural interpretation is that the dynamics are influenced by a non-negligible interaction between the diagonal neighbors. The expected interaction strength for two atoms at a distance of \SI{9.9}{\micro\meter} with an angle of \SI{45}{\degree} between the interatomic and quantization axis amounts to ${|\Delta E|/h =\SI{75}{\kilo\hertz}}$. These interactions gain further importance as the number of neighbors with strong coupling to the Rydberg state varies within the structure. In particular for the center sites with four neighbors with high probability of excitation, this non-negligible energy shift leads to a reduced amplitude of the oscillations, although ${\Omega>|\Delta E|/\hbar}$. A fluctuating number of interacting atoms between experimental realizations adds to dephasing. This is plausible, as the preparation efficiency of the checkerboard cluster is finite and the data were not post-selected for a fixed number of neighbors.

\subsection*{Collective Rabi oscillations}
The Rydberg blockade effect could be verified through the observation of a collective enhancement of the Rabi oscillation between the ground state and an entangled state with a single Rydberg excitation shared among two and three atoms, respectively.
The experiments, as presented in Fig. 5, were facilitated by increasing the interaction strength with the use of the higher lying Rydberg state $\mathrm{|87D_{5/2},m_J=-3/2\rangle}$ as well as by the repeated assembly of optimized target structures. For the obtained dynamics, the above discussion of experimental limitations applies as well. The depicted signals are corrected for static effects, including the  Rydberg detection error of ${\SI{0.06}{}~(n=\SI{87}{})}$. Even though the damping of the oscillation becomes more pronounced when more than one atom is present, the enhancement of the Rabi frequency is clearly visible and in good agreement with the expected scaling with $\sqrt{N}$ (see Tab.~\ref{tab:2}).\\
There appears to be an imperfect blockade in the triangular geometry of three atoms for longer pulse durations. Analysis of the residual double excitation probability reveals it to be higher than expected from calculated blockade strengths. This points to either an overestimation of the van der Waals strength or to a diminution of the blockade through three-body interactions \cite{Pohl2009,Qian2013}. These results underpin the notion that, in order to rely on a robust blockade mechanism in an array of Rydberg atoms, one has to carefully tune the experimental parameters when working with $\mathrm{D}$ states. For 2D geometries, switching to $\mathrm{S}$ states providing robust and isotropic blockade interaction may be preferable.

\subsection*{Future improvements}
The analysis of the observed dynamics and experimental limitations leads to the identification of the parameters with the highest potential for improvement. Increasing the Rabi frequency and the intermediate state detuning by an appreciable amount would render the system less vulnerable to incoherent scattering and Stark-shift induced fluctuations in two-photon detuning. Doppler shift fluctuations could be reduced by a factor of three by changing the excitation geometry to one utilizing an anti-parallel alignment of the Rydberg laser beams. The implementation of these measures will allow us to access the full potential of Rydberg-coupled quantum systems in large-scale MLA-based single-atom platforms.

\section{Conclusion}

We have presented the first realization of Rydberg-interacting systems in microlens-generated single-atom arrays. Atom-by-atom assembly of defect-free atom patterns enabled the site-resolved characterization of coherent Rydberg coupling and the observation of the Rydberg blockade effect. A detailed analysis of the experimental limitations provides a clear road towards improved quantum state control. Benefitting from the versatility and scalability of the micro-optical approach, the presented platform lends itself to prospect applications in quantum information sciences, such as quantum computation, simulation, and sensing \cite{Dumke2016}.

\ack
We acknowledge financial support by the Deutsche Forschungsgemeinschaft (DFG) [Grant No. BI 647/6-1 and BI 647/6-2, Priority Program SPP 1929 (GiRyd)]. We thank the \textsc{labscript suite} \cite{Starkey2013} community for support in implementing state-of-the-art control software for our experiments.

\section*{References}
\bibliographystyle{unsrt}
\bibliography{RydbergArray}

\end{document}